\documentclass[aps,prd,preprint,superscriptaddress]{revtex4-1}

\usepackage{amsmath}
\usepackage{dcolumn}
\usepackage{bm}

\begin{document}

\title{Scattering in pseudo-Hermitian quantum field theory and causality violation}

\author{Oleg O. Novikov}
\email[]{o.novikov@spbu.ru}
\affiliation{Saint Petersburg State University, 7/9 Universitetskaya nab.,
St. Petersburg, 199034, Russia}

\date{\today}

\begin{abstract}
The non-Hermitian but $\mathcal{PT}$-symmetric quantum field theories are known
to have a pseudo-Hermitian interpretation. However the corresponding
intertwining operator happens to be nonlocal that raises the question to
what extent this nonlocality affects observable quantities. We consider
the case when the intrinsic parity of the interaction terms is determined by degree
of coupling constant. We show that the perturbative S-matrix of the equivalent
Hermitian description can be easily obtained from the perturbative S-matrix
of the non-Hermitian model. Namely, the first order vanishes whereas
the second order is given by the real part of the second order T-matrix
of the non-Hermitian model. We compute directly the 2-point and 4-point
correlation functions in the equivalent Hermitian model for the $i\phi^3$
model and find the results consistent with this relation. The 1-loop correction
to the mass happens to be real reflecting the disappearance of 2-body decays.
However the 2 to 2 scattering amplitude obtained using LSZ formula has poles
taken in principal value which implies the violation of the causality.
\end{abstract}


\maketitle

\section{Introduction}

The non-Hermitian $\mathcal{PT}$-symmetric quantum theories
have attracted significant attention due to their unusual properties
\cite{BenderRealSpectra,AndrianovRealSpectra,BenderReview}.
Surprisingly they often possess purely real spectrum and produce
an unitary evolution with respect to the non-standard inner product. As a
matter of fact these Hamiltonians may be shown to be connected
to the Hermitian Hamiltonians through the non-unitary intertwining
operator and thus represent a particular class of the so-called
pseudo-Hermitian Hamiltonians
\cite{MostafazadehPH}.
It is also interesting that some
seemingly unstable Hamiltonians may be reinterpreted as
non-Hermitian $\mathcal{PT}$-symmetric Hamiltonians and
then as pseudo-Hermitian Hamiltonians with purely real positive
spectrum.

While most of the studies concentrated on the finite-dimensional
quantum mechanical models the $\mathcal{PT}$-symmetric quantum
field theories were also investigated \cite{BenderSD1,BenderPhi3_1,
BenderPhi3_2,BenderCScalar,
BenderSD2,Alexandre1,Alexandre2,Alexandre3}.
This line of research is
especially important because it gives a hope that some models
interesting from a phenomenological perspective but plagued by
unitarity and stability problems may be reinterpreted as consistent
quantum field theories \cite{BenderLee,BenderUnstable}.
It inspired a number of proposals for the particle physics and
cosmology \cite{AlexandreN1,AlexandreN2,OhlssonN,MannheimCG,
AndrianovPTom1,AndrianovPTom2,AndrianovSpringer,
NovikovPEPAN,NovikovEPJWC,Braun1,Braun2}.

However many basic questions about the $\mathcal{PT}$-symmetric
quantum field theories that are important for their applicability have
not yet been sufficiently investigated. One of such questions is whether
these quantum field theories satisfy the relativistic causality.
There is a very good reason to suspect that despite relativistic invariance
this may not be a case. As was shown in \cite{BenderPhi3_1,BenderPhi3_2,
BenderCScalar} for the $i\phi^3$ quantum field theory the intertwining
operator is nonlocal,
\begin{equation}
Q\simeq\int d^3x d^3y d^3z
\Big[\mathcal{M}(\vec{x},\vec{y},\vec{z})\pi(x)\pi(y)\pi(z)
+\mathcal{N}(\vec{x}|\vec{y},\vec{z})\phi(y)\pi(x)\phi(z)
\Big],
\label{QNonlocal}
\end{equation}
As result the equivalent Hermitian Hamiltonian given by
$h=e^{-Q/2}H e^{Q/2}$ is also nonlocal.
This does not affect the locality of the vacuum correlators
of the fields of the non-Hermitian description that were extensively
studied in the literature \cite{BenderSD1,BenderSD2}. However because
such fields are not Hermitian under the modified norm they are related to
the observable fields through the same nonlocal intertwining operators. Therefore
one may expect the appearance of the causality violations.

In this paper we are studying the S-matrix in a non-Hermitian
$\mathcal{PT}$-symmetric quantum field theory interpreted in a
pseudo-Hermitian fashion. We present a very simple relation between S-matrices
in the non-Hermitian and equivalent Hermitian descriptions. However this simple
result shows the generic violation of the causality when the initial non-Hermitian
model is local. We confirm this by directly computing the propagator and 2 to 2
scattering amplitude in the equivalent Hermitian model.

It should be noted that the issue of nonlocality of
the $\mathcal{PT}$-symmetric quantum models with local non-Hermitian
Hamiltonians arises already in the context of the finite dimensional quantum
mechanical systems. Its influence on the scattering and the possibility to
save the causality by relaxing the locality of the non-Hermitian Hamiltonian was
studied in \cite{JonesScatter1,ZnojilScatter1,ZnojilScatter2,ZnojilScatter3}.
To our knowledge our paper is the first one to address these questions in the
context of the $\mathcal{PT}$-symmetric QFT.

The paper is organized as follows. In Section \ref{PTintro} we briefly
elucidate the basics of the pseudo-Hermitian models. In Section
\ref{EtaPerturb} we review the perturbation theory for the intertwining operator.
In Section \ref{FormalS} this is used to compute the formal perturbative
S-matrix of the equivalent Hermitian model. In Section \ref{QinQFT} we
derive the intertwining operator for the generic local $\mathcal{PT}$-symmetric
QFT in the momentum representation using a method that in our opinion is
significantly simpler compared to the previous work. Section \ref{SinQFT}
concludes this computation of the formal S-matrix by showing that the commutator
between the intertwining operators at large times vanishes. This leads to the
extremely simple relation for the S-matrices of two equivalent descriptions
however we also demonstrate the violation of the Bogolyubov microcausality.
In Section \ref{PTQFT} we formulate the simplest non-trivial
$\mathcal{PT}$-symmetric QFT and in Section \ref{Correlators} we
develop a technique to compute correlation functions of the observable
fields of the Hermitian model equivalent to this QFT. In Sections
\ref{Propagator} and \ref{ScatteringAmplitude} we apply this technique
to the propagator and 2 to 2 scattering amplitude. In Section
\ref{NoHermiticity} we relax the assumption of the Hermiticity of the
intertwining operator in an attempt to restore the causality. In Conclusions
we summarize the results and discuss the prospects.

\section{$\mathcal{PT}$-symmetric quantum theory\label{PTintro}}

Consider the non-Hermitian but $\mathcal{PT}$ Hamiltonian,
\begin{equation}
H\neq H^\dagger,\quad [\mathcal{PT},H]=0,
\end{equation}
where $\mathcal{P}$ is the \textit{intrinsic parity} operator (that reflects
only fields and not spatial coordinates) and $\mathcal{T}$ is the usual time
reflection operator. They are defined on the canonical fields $\phi(\vec{x})$
and their momenta $\pi(\vec{x})$ in the following way,
\begin{align}
\mathcal{P}:\quad&\phi(\vec{x})\mapsto -\phi(\vec{x}),\quad&
\pi(\vec{x})\mapsto-\pi(\vec{x}),\quad&i\mapsto i\\
\mathcal{T}:\quad&\phi(\vec{x})\mapsto \phi(\vec{x}),\quad&
\pi(\vec{x})\mapsto-\pi(\vec{x}),\quad&i\mapsto -i,
\end{align}
where the action on the imaginary unit means that $\mathcal{P}$ is
a linear operator whereas $\mathcal{T}$ is antilinear. If one introduces
the creation and annihilation operators,
\begin{align}
\phi(\vec{x})=\int\frac{d^3k}{(2\pi)^3} \frac{1}{\sqrt{2E_{\vec{k}}}}
\left(a_{\vec{k}}^\dagger e^{-i\vec{k}\vec{x}}+
a_{\vec{k}} e^{i\vec{k}\vec{x}}\right),\\
\pi(\vec{x})=\int\frac{d^3k}{(2\pi)^3} i\sqrt{\frac{E_{\vec{k}}}{2}}
\left(a_{\vec{k}}^\dagger e^{-i\vec{k}\vec{x}}-
a_{\vec{k}} e^{i\vec{k}\vec{x}}\right),\label{picanonical}
\end{align}
the definitions above result in the following action,
\begin{equation}
\mathcal{P}a_{\vec{k}}\mathcal{P}=-a_{\vec{k}},\quad
\mathcal{T}a_{\vec{k}}\mathcal{T}=a_{-\vec{k}},
\end{equation}

As was shown in \cite{BenderRealSpectra} such Hamiltonians often have purely
real spectrum. This allows to interpret them as a pseudo-Hermitian ones,
i.e. related to some Hermitian Hamiltonian with the intertwining operator
\cite{MostafazadehPH},
\begin{equation}
h\equiv \eta H \eta^{-1},\quad
H^\dagger \eta^\dagger\eta=\eta^\dagger \eta H.
\label{PHdef}
\end{equation}
The latter equation guarantees the Hermiticity of $h$ with respect to
the initial product or equivalently the Hermiticity of $H$ with respect to
the new inner product,
\begin{equation}
(\Psi,\Phi)\equiv \langle\Psi|\eta^\dagger\eta|\Phi\rangle.
\label{innerProd}
\end{equation}
We stress that this inner product is positively definite by construction and
thus no negative norm states are needed. Note that the initial variable
$\phi$ is not Hermitian under this inner product
and thus is not observable. To get the observable that would correspond to
the $\phi$ in the equivalent Hermitian description one has to consider
$\eta^{-1}\phi\eta$. In conclusion, there are two alternative description
of the same model that are summarized in the Table \ref{tab:table1}. One
may note the tradeoff between simplicity of the observables and simplicity
of the evolution.
\begin{table}[b]
\caption{\label{tab:table1}
The correspondence between the objects in the description with the
non-Hermitian Hamiltonian and the equivalent Hermitian one
}
\begin{ruledtabular}
\begin{tabular}{lcc}
\textrm{Object}&
\textrm{Non-Hermitian}&
\textrm{Equivalent Hermitian}\\
\colrule
Hamiltonian & $H$ & $h=\eta H\eta^{-1}$\\
Inner product & $(\Psi,\Phi)=\langle\Psi|\eta^\dagger\eta|\Phi\rangle$
 & $\langle\Psi|\Phi\rangle$\\
Unobservable field & $\phi$ & $\eta\phi\eta^{-1}$\\
Observable field & $\eta^{-1}\phi\eta$ & $\phi$\\
\end{tabular}
\end{ruledtabular}
\end{table}

The relations above were written for time-independent $H$ in the
Schr{\"o}dinger picture. In the time-dependent case
\cite{MostafazadehNonstat,CannataNonstat,FringMoussa,ZnojilTD1,ZnojilTD2}
the relation (\ref{PHdef}) is no longer true and extra non-Hermitian term
appears. However one can always define $\eta$ through the following property,
\begin{equation}
\eta(t_2)^\dagger\eta(t_2)U_H(t_2,t_1)=
\Big[U_H(t_2,t_1)^\dagger\Big]^{-1}\eta(t_1)^\dagger\eta(t_1).
\label{PHdef2}
\end{equation}
where $U_H$ is the evolution operator generated by $H$. This relation
simply means that the inner product $(\Psi,\Phi)$ is conserved by the
temporal evolution of the state vectors and holds in all pictures.

Though this is not required for the hermiticity of $h$ one usually
assumes that the operator $\eta$ transforms in the following way
so that the $\mathcal{PT}$ symmetry of $H$ was conserved also
by $h$,
\begin{equation}
\mathcal{P}\eta^\dagger\mathcal{P}=\eta^{-1},\quad
\mathcal{T}\eta^\dagger\mathcal{T}=\eta^{-1}.
\label{etaExtra}
\end{equation}
Also one usually fixes $\eta$ to be Hermitian thus fixing the arbitrary
unitary transformation one can insert into the intertwining operator.
In this paper we will hold to the assumptions of the Hermiticity and
$\mathcal{P}$ parity but \emph{will not} hold to the assumption of
the $\mathcal{T}$ parity.

For the restricted class of Hamiltonians
$\mathcal{P}H\mathcal{P}=H^\dagger$ the operator
$\mathcal{C}$ is often introduced 
that have the following properties,
\begin{equation}
\mathcal{C}\equiv \mathcal{P}\eta^\dagger\eta,\,
[\mathcal{C},H]=0,\, [\mathcal{C},\mathcal{PT}]=0,\,
\mathcal{C}^2=1,
\end{equation}
where all (\ref{etaExtra}) are implied. Despite its notation it is not
related to the charge conjugation operator. The useful property
of this operator is that $\mathcal{C}$ of an eigenstate of $H$
coincides with its $\mathcal{PT}$ parity. Thus if the e.g. the
vacuum state $|\Omega\rangle$ is $\mathcal{PT}$ symmetric
then $\mathcal{C}|\Omega\rangle=1$ that significantly simplifies
the computations on the vacuum state \cite{JonesCop}.

\section{Perturbation theory for $\eta$\label{EtaPerturb}}

Let us now assume that $H$ can be represented in the following form,
\begin{align}
H=\sum_{k=0}^{+\infty}(ig)^k H_k,\quad H_k=H_k^\dagger,
\quad
\mathcal{P}H_k\mathcal{P}=(-1)^k H_k,\quad
\mathcal{T}H_k\mathcal{T}=H_k,\label{Hseries}
\end{align}
where the coupling constant $g$ is assumed to be small. For purely
imaginary coupling constants the operator becomes Hermitian,
\begin{equation}
\tilde{H}=\sum_{k=0}^{+\infty}g^k H_k.
\end{equation}
We will assume that the corresponding evolution operators are also
related by such analytic continuation at least within perturbation theory.
I.e. if,
\begin{equation}
U_{\tilde{H}}(t_2,t_1)=\mathrm{T}
\exp\left[-i\int_{t_1}^{t_2}dt \tilde{H}\right]
=\sum_{k=0}^{+\infty}g^k U_{H,k},
\end{equation}
we will assume that,
\begin{equation}
U_H(t_2,t_1)=\sum_{k=0}^{+\infty}(ig)^k U_{H,k}
\end{equation}

We apply the similar decomposition to the equivalent Hermitian
operator $h$ and to the intertwining operator $\eta$ (we omit the usual factor
$1/2$ to simplify the equations),
\begin{equation}
h=\sum_{k=0}^{+\infty}g^k h_k,\quad
\eta=\exp\left[-\sum_{k=0}^{+\infty}\frac{g^{2k+1}}{(2k+1)!}Q_k\right],
\end{equation}
where we omit the us $Q_k$ is assumed to have the following properties so that $\eta$
satisfied extra assumptions of hermiticity and (\ref{etaExtra}),
\begin{equation}
Q_k^\dagger=Q_k,\quad \{Q_k,\mathcal{P}\}=0.
\label{Qextra}
\end{equation}
As with $\eta$ one may also demand,
\begin{equation}
\{Q_k,\mathcal{T}\}=0.
\label{QextraT}
\end{equation}
This make it possible to choose the ansatz with only odd powers of $g$.
As we will restrict ourselves only to the computations up to the second order
in $g$ we will denote $Q\equiv Q_0$.

Then from (\ref{PHdef}) one obtains the following results \cite{BenderReview},
\begin{equation}
i[H_0,Q]=H_1,\label{Qeq1}
\end{equation}
\begin{equation}
h_1=0,\quad h_2=-H_2-\frac{i}{2}[Q,H_1].\label{EquivH}
\end{equation}

To use (\ref{Qeq1}) we note the following fact. As we work in the
perturbation theory let us go into the interaction picture,
\begin{align}
Q(t)=e^{iH_0t}Q e^{-iH_0t},\quad
H_k(t)=e^{iH_0t}H_k e^{-iH_0t}.
\end{align}
Then (\ref{Qeq1}) becomes a very simple relation,
\begin{equation}
\partial_t Q(t)= H_1(t),\quad
Q(t_2)-Q(t_1)=\int_{t_1}^{t_2}dt\, H_1(t).
\label{Qformula}
\end{equation}
We would like to stress that this relation can be obtained from the
general equation for (\ref{PHdef2}) written in the interaction picture.
Thus this results holds even if $H_1$ depends on $t$ explicitly, e.g.
if one turns off interaction asymptotically.

No matter how simple (\ref{Qformula}) may appear it has an important
consequence. Even if the interaction falls down at large times generally
speaking the intertwining operator remains to be nontrivial. Thus even
though $H$ may appear Hermitian asymptotically, the field variables of
the non-Hermitian theory do not become observables and do not create
good asymptotic particle states that are orthogonal with each other.

Thus to understand the actual dynamics of the $\mathcal{PT}$-symmetric
QFT one has to use either the non-Hermitian Hamiltonian $H$ and
the modified field variables $\eta\phi(x)\eta^{-1}$ or equivalently
the Hermitian Hamiltonian $h$ and usual field variables. These two approaches
should be fully equivalent but in this paper we choose the second one.

\section{Formal $S$-matrix\label{FormalS}}

Let us consider the evolution operator of the equivalent Hermitian model
in the interaction picture,
\begin{equation}
U_h^{(I)}(t_f,t_0)=e^{iH_0t_f}e^{-ih(t_f-t_0)}e^{-iH_0t_0}.
\end{equation}
We can rewrite it in terms of the similar operator for the non-Hermitian
Hamiltonian,
\begin{align}
U_h^{(I)}(t_f,t_0)=e^{iH_0t_f} \eta e^{-iH(t_f-t_0)} \eta^{-1} e^{-iH_0t_0}
=
\eta(t_f) U_H^{(I)} \eta(t_0)^{-1},
\end{align}
where we define,
\begin{equation}
\eta(t)\equiv e^{iH_0t}\eta e^{-iH_0t}\simeq e^{-gQ(t)}+\mathcal{O}(g^3)
\end{equation}

Let us now decompose the evolution operators into series in $g$,
\begin{equation}
U_h^{(I)}\simeq 1+gU_{h,1}^{(I)}+g^2U_{h,2}^{(I)}+\mathcal{O}(g^3)
\end{equation}
\begin{equation}
U_H^{(I)}\simeq 1+igU_{H,1}^{(I)}-g^2U_{H,2}^{(I)}+\mathcal{O}(g^3)
\end{equation}
As noted above we assume that the perturbation series of $U_H^{(I)}$ can be
obtained by analytical continuation in $g$ of the perturbation series of the
unitary operator $U_{\tilde{H}}^{(I)}$. Then the unitarity implies,
\begin{equation}
U_{H,1}=-(U_{H,1})^\dagger,\quad
 2\Re\Big[U_{H,2}\Big]=\Big[U_{H,1}\Big]^2.\label{unitarity}
\end{equation}
Using the Dyson expansion,
\begin{equation}
U_{H,1}^{(I)}(t_f,t_0)=-i\int_{t_0}^{t_f}dt\, H_1(t),\label{Dyson1}
\end{equation}
and the relation (\ref{Qformula}) one finds,
\begin{equation}
U_{h,1}^{(I)}=-Q(t_f)+Q(t_0)+\int_{t_0}^{t_f}dt\, H_1(t)=0.
\end{equation}
This should not come as surprise because as we have noted in (\ref{EquivH})
the first order of $h$ vanishes.

For the second order we get,
\begin{align}
U_{h,2}^{(I)}=-U_{H,2}^{(I)}+\frac{Q^2(t_f)}{2}+\frac{Q^2(t_0)}{2}
-Q(t_f)Q(t_0)
-iQ(t_f)U_{H,1}^{(I)}+iU_{H,1}^{(I)}Q(t_0)
\end{align}
Again using (\ref{Dyson1}) and (\ref{Qformula}) we rewrite it as,
\begin{equation}
U_{h,2}^{(I)}=-U_{H,2}^{(I)}-\frac{1}{2}\Big[U_{H,1}^{(I)}\Big]^2
+\frac{1}{2}[Q(t_f),Q(t_0)].
\end{equation}
Finally using (\ref{unitarity}) we see that the second term cancels the real
part of the first one,
\begin{equation}
U_{h,2}^{(I)}=-\Im\Big[U_{H,2}^{(I)}\Big]+\frac{1}{2}[Q(t_f),Q(t_0)].
\label{U2}
\end{equation}

As usual we apply the practical definition of the S-matrix as the evolution
operator in the interaction picture in the limit of large times (assuming
that such limit exists),
\begin{align}
S_h\equiv \lim_{\substack{t_f\rightarrow+\infty\\t_0\rightarrow -\infty}}
U_h^{(I)}(t_f,t_0),\label{Shmatrix}\\
S_H\equiv \lim_{\substack{t_f\rightarrow+\infty\\t_0\rightarrow -\infty}}
U_H^{(I)}(t_f,t_0),
\end{align}
Again assuming the validity of the validity of the analytical continuation
from the unitary S-matrix to $S_H$ we introduce the T-matrix series in powers of $g$,
\begin{align}
S_h\simeq 1+ig T_h^{(1)}+ig T_h^{(2)}+\mathcal{O}(g^3),\\
S_H\simeq 1-g T_H^{(1)}-ig T_H^{(2)}+\mathcal{O}(g^3).
\end{align}

Let us introduce the asymptotic intertwining operators,
\begin{equation}
Q_{in}=\lim_{t\rightarrow -\infty}Q(t),\quad
Q_{out}=\lim_{t\rightarrow +\infty}Q(t).
\end{equation}

As the limit of (\ref{U2}) we obtain,
\begin{equation}
T_h^{(2)}=-\Re\Big[T_H^{(2)}\Big]-\frac{i}{2}[Q_{out},Q_{in}].
\label{Tmatrix1}
\end{equation}

\section{Intertwining operator in QFT\label{QinQFT}}

Let $H$ describe a $\mathcal{PT}$-symmetric quantum field theory
obtained through the analytical continuation in $g$ of the local
quantum field theory.
\begin{equation}
H_1(t)=\int d^3x V_1(t,\vec{x}),\,
\mathcal{P}V_1\mathcal{P}=-V_1,
\,\mathcal{T}V_1\mathcal{T}=-V_1.
\label{localH1}
\end{equation}
Then in the perturbation theory one should
be able to represent the interaction terms as combinations of the
creation and annihilation operators,
\begin{equation}
H_1(t)=\sum_{\{\varepsilon_k\}}\int \prod_k
\frac{d^3p_k}{E_{\vec{p}_k}}
\mathcal{V}_{\{\varepsilon_k\}}(\{\vec{p}_k\})
e^{i\sum_k\varepsilon_k E_{\vec{p_k}}t}
\delta^{(3)}\Bigg(\sum_k\varepsilon_k \vec{p_k}\Bigg),
\label{Vdecomp}
\end{equation}
where $\mathcal{V}$ is some operator valued distribution constructed
as some $c$-function of $\{\varepsilon_k\}$ and $\{\vec{p}_k\}$ multiplied
on a combination of the creation and annihilation operators in accordance
with multiindex $\{\varepsilon_k\}$ so that $\varepsilon_k=+1$
corresponds to $a_{\vec{p}_k}^\dagger$ an $\varepsilon_k=-1$
corresponds to $a_{\vec{p}_k}$. The negative $\mathcal{P}$-parity
means that only terms with odd number of these operators contribute.

The hermiticity $H_1^\dagger=H_1$ means that,
\begin{equation}
\mathcal{V}_{\{\varepsilon_k\}}(\{\vec{p}_k\})=
\bar{\mathcal{V}}_{\sigma\{-\varepsilon_k\}}(\sigma\{\vec{p}_k\}),
\label{Vhermiticity}
\end{equation}
and $\mathcal{T}H_1\mathcal{T}=H_1$ yields,
\begin{equation}
\mathcal{V}_{\{\varepsilon_k\}}(\{\vec{p}_k\})=
\bar{\mathcal{V}}_{\{\varepsilon_k\}}(\{-\vec{p}_k\}),
\label{VTsymmetry}
\end{equation}
where $\bar{\mathcal{V}}$ denotes the complex conjugation of
the $c$-numerical coeffiecients and $\sigma$ reverses the order in
the multiindex.

We assume that a similar decomposition may be written for $Q(t)$,
\begin{equation}
Q(t)=\sum_{\{\varepsilon_k\}}\int \prod_k
\frac{d^3p_k}{E_{\vec{p}_k}}
\mathcal{Q}_{\{\varepsilon_k\}}(\{\vec{p}_k\})
e^{i\sum_k\varepsilon_k E_{\vec{p_k}}t}
\delta^{(3)}\Bigg(\sum_k\varepsilon_k \vec{p_k}\Bigg).
\end{equation}

Then (\ref{Qformula}) yields the distribution equation,
\begin{equation}
i\Big[\sum_k\varepsilon_k E_{\vec{p_k}}\Big]
\mathcal{Q}_{\{\varepsilon_k\}}(\{\vec{p}_k\})
=\mathcal{V}_{\{\varepsilon_k\}}(\{\vec{p}_k\}),
\end{equation}
that has a general solution,
\begin{align}
\mathcal{Q}_{\{\varepsilon_k\}}(\{\vec{p}_k\})=
-\mathcal{V}_{\{\varepsilon_k\}}(\{\vec{p}_k\})
\mathrm{P.v.}\frac{i}{\sum_k\varepsilon_k E_{\vec{p_k}}}
+\mathcal{A}_{\{\varepsilon_k\}}(\{\vec{p}_k\})
\delta(\sum_k\varepsilon_k E_{\vec{p_k}})
\end{align}
where $\mathrm{P.v.}$ denotes the principal value and $\mathcal{A}$
is an arbitrary Hermitian operator value distribution. This reflects the
freedom to add an arbitrary operator commuting with $H_0$. The
most significant constraint comes from the Lorentz invariance of the inner
product. We do not derive the
most general form of $Q$ but instead simply choose,
\begin{equation}
\mathcal{A}_{\{\varepsilon_k\}}(\{\vec{p}_k\})=\alpha
\mathcal{V}_{\{\varepsilon_k\}}(\{\vec{p}_k\}).
\end{equation}
This simple ansatz results in,
\begin{align}
Q(t)=&-i\sum_{\{\varepsilon_k\}}\int \prod_k
\frac{d^3p_k}{E_{\vec{p}_k}}
\mathcal{V}_{\{\varepsilon_k\}}(\{\vec{p}_k\})
\nonumber\\&
\left[
\mathrm{P.v.}\frac{e^{i\sum_k\varepsilon_k E_{\vec{p_k}}t}}
{\sum_k\varepsilon_k E_{\vec{p_k}}}
\delta^{(3)}\Bigg(\sum_k\varepsilon_k \vec{p_k}\Bigg)
+i\alpha\delta^{(4)}\Bigg(\sum_k\varepsilon_k p_k\Bigg)\right].
\label{Qfinal}
\end{align}
Such $Q$ obviously has the same $\mathcal{P}$-parity as $H_1$.
From (\ref{Vhermiticity}) and evenness of the $\delta$-function
follows that the hermiticity of $Q$ requires $\alpha=\alpha^\ast$.
This is exactly what we need if we want to shift the pole in the fraction
in the complex plane. However (\ref{VTsymmetry}) and symmetry
of the second term under reflection of $\vec{p}_k$ means that
for (\ref{QextraT}) to be true one needs $\alpha=-\alpha^\ast$.
Thus to shift the pole one has to relax this extra assumption
and thus break $\mathcal{PT}$-symmetry of $h$.

One may check that (\ref{Qfinal}) indeed conserves the Lorentz invariance
of the inner product (\ref{innerProd}). The operator of the Lorentz
boost $\Lambda$ characterized by rapidity $\vec{\beta}$ may be written as,
\begin{equation}
\mathcal{U}_g(\Lambda)\simeq 1+i\beta_i \Big( L_0^{0i}(t)
 +i g L_0^{0i}(t)\Big)
\end{equation}
where $L_0^{0i}$ is the boost operator in the free QFT and,
\begin{align}
L_1^{0i}(t)&=-\int d^3x\, x^iV_1(t,\vec{x})
\nonumber\\&
=i\sum_{\{\varepsilon_k\}}\int \prod_k
\frac{d^3p_k}{E_{\vec{p}_k}}
\nabla_i^{(P)}\left[
\mathcal{V}_{\{\varepsilon_k\}}(\{\vec{p}_k\})
\right]
e^{i\sum_k\varepsilon_k E_{\vec{p_k}}t}
\delta^{(3)}\Bigg(\sum_k\varepsilon_k \vec{p_k}\Bigg),
\label{VBoost}
\end{align}
where we define the total momentum as,
\begin{equation}
\vec{P}=\sum_k\varepsilon_k \vec{p}_k,
\quad
P^0=\sum_k\varepsilon_k E_{\vec{p_k}}.
\end{equation}

We assume that $V_1(t,\vec{x})$ represented using interaction picture fields
transforms as a Lorentz scalar field with respect to the boost of the
\emph{free} QFT.
As in (\ref{Vdecomp}) we used the Lorentz
invariant integration measure $\mathcal{V}$ transforms as a Lorentz scalar too,
\begin{equation}
\Big[\mathcal{U}_0(\Lambda)\Big]^{-1}
\mathcal{V}_{\{\varepsilon_k\}}(\{\vec{p}_k\})\mathcal{U}_0(\Lambda)
=\mathcal{V}_{\{\varepsilon_k\}}(\{\Lambda^{-1}\vec{p}_k\})\vert_{p_k^0
=E_{\vec{p}_k}},
\end{equation}
For an infinitesimal boost
$\Lambda^{-1}\vec{p}_k\simeq \vec{p}_k-\vec{\beta}E_{\vec{p}_k}$ this implies,
\begin{equation}
[L_0^{0i}(t),\mathcal{V}_{\{\varepsilon_k\}}(\{\vec{p}_k\})]=
i P^0\nabla_i^{(P)}
\left[\mathcal{V}_{\{\varepsilon_k\}}(\{\vec{p}_k\})\right].
\end{equation}
Combining this with (\ref{Qfinal}) and (\ref{VBoost}) one easily obtains that the
inner product (\ref{innerProd})
is indeed Lorentz invariant,
\begin{equation}
\Big[\mathcal{U}(\Lambda)\Big]^\dagger
\eta^\dagger\eta\mathcal{U}(\Lambda)-\eta^\dagger\eta
\simeq
2ig\beta_i\Big([L_0^{0i}(t),Q(t)]+iL_1^{0i}(t)\Big)=0,
\end{equation}

This confirms and generalizes the result obtained in \cite{BenderCScalar}.
Indeed in case of the $i\phi^3$ model if $\alpha=0$ (\ref{Qfinal}) gives
the same nonlocal intertwining operator (\ref{QNonlocal}) as the one
studied in \cite{BenderPhi3_1,BenderPhi3_2,BenderCScalar}.

\section{Formal $S$-matrix in $\mathcal{PT}$-symmetric QFT\label{SinQFT}}

To find $Q_{out}$ and $Q_{in}$ we first explicitly compute $T_{H}^{(1)}$
by integrating (\ref{Vdecomp}) and using the standard integral representation
of the $\delta$-function,
\begin{align}
T_{H}^{(1)}=2\pi
\sum_{\{\varepsilon_k\}}\int \prod_k
\frac{d^3p_k}{E_{\vec{p}_k}}
\mathcal{V}_{\{\varepsilon_k\}}(\{\vec{p}_k\})
\delta^{(4)}\Bigg(\sum_k\varepsilon_k p_k\Bigg),
\quad p_k^0\equiv E_{\vec{p}_k},
\end{align}
we use the following identity \cite{Schweber} in a sense
of a distribution on localized wavepackets in $E$ (that usually represented
by shifting $t_f\mapsto t_f (1-i\epsilon)$ in the QFT textbooks),
\begin{equation}
\mathrm{P.v.}\frac{e^{iEt}}{E}=\frac{1}{2}\frac{e^{iEt}}{E+i\epsilon}
+\frac{1}{2}\frac{e^{iEt}}{E-i\epsilon}
\xrightarrow[t\rightarrow\pm\infty]{} \pm\pi i\delta(E)
\end{equation}

This yields,
\begin{align}
Q_{out}=\frac{\pi+\alpha}{2\pi}T_H^{(1)},\quad
Q_{in}=\frac{-\pi+\alpha}{2\pi}T_H^{(1)}
\end{align}
This result makes the violation of $\mathcal{T}$-symmetry by nonzero
$\alpha$ easily recognizable. One can also easily see that,
\begin{equation}
[Q_{out},Q_{in}]=0,
\end{equation}
and thus (\ref{Tmatrix1}) simplifies to,
\begin{equation}
T_h^{(2)}=-\Re\Big[T_H^{(2)}\Big].
\label{FinalT}
\end{equation}

One may then easily show that if $H_1$ is the local Hamiltonian 
the S-matrix of the equivalent Hermitian theory
will almost always lead to the causality violation.
E.g. the Bogolyubov microcausality condition requires \cite{Bogolubov},
\begin{equation}
\frac{\delta}{\delta\phi_k(z)}
\Bigg(\frac{\delta S}{\delta\phi_j(y)}S^\dagger\Bigg)
\sim \theta(z^0-y^0)\theta\Big((z-y)^2\Big),
\label{microcausality}
\end{equation}
This condition means that if we represent the evolution of the
wavefunctional as a sequence of the scattering events of the
localized wavepackets, the secondary wavepackets will be
produced only in the future lightcones of regions of the intersections
of the primary wavepackets. This guarantees that both ordinary
causality (future does not influence the past) and the relativistic
causality (no superluminal propagation) holds.

If $H_1$ is given by (\ref{localH1}) and $H_2$ is also a local
operator represented as a integral of $V_2(x)$ then using
(\ref{FinalT}) we obtain,
\begin{align}
T_h^{(2)}=-\int d^4x V_2(x)
+
\frac{i}{4}\int d^4x_1 d^4x_2
\varepsilon(x_1^0-x_2^0)[V_1(x_1),V_2(x_2)],
\end{align}
Then we get,
\begin{align}
\frac{\delta}{\delta\phi_k(z)}
\Bigg(\frac{\delta S_h}{\delta\phi_j(y)}S_h^\dagger\Bigg)
\simeq
\frac{g^2}{2}\varepsilon(y^0-z^0)
\Big[\frac{\partial V_1}{\partial\phi_j}(y),
\frac{\partial V_1}{\partial\phi_k}(z)\Big],
\label{CausalityViolation}
\end{align}
that is nonzero not only in the future lightcone but also
in the past lightcone. Therefore the secondary wavepackets
are produced in both the future and the past lightcones of the intersection
regions of the primary wavepackets. This obviously violates the ordinary
causality. In a sequence of collisions this also leads to the superluminal
propagation however there is possibility that this violation of the relativistic
causality is somehow compensated in the higher orders of the S-matrix.
The obvious exception is the linear potential $V_1(x)=a\phi(x)$. Then the
operator $\eta$ becomes simply a shift of the field variable in the imaginary
direction that may be reabsorbed into other interactions without violation
of locality.

\section{Example of the $\mathcal{PT}$-symmetric QFT\label{PTQFT}}

Let us consider the model with the following action,
\begin{align}
S=\int d^4x\Bigg[\frac{1}{2}\partial_\mu\phi_k\partial^\mu\phi_k
-\frac{m_k^2}{2}(\phi_k)^2
-ig\frac{\mu_{ijk}}{3!}\phi_i\phi_j\phi_k
-g^2\frac{\lambda_{ijkl}}{4!}\phi_i\phi_j\phi_k\phi_l\Bigg],
\end{align}
where all coupling constants are assumed to be real and totally symmetric
under permutations of indices. As usual we use the canonical momenta
of the free theory $\pi_k(\vec{x})\equiv \dot{\phi}_k(\vec{x})$ that may
be represented as (\ref{picanonical}). Then the first few terms of the
Hamiltonian in the (\ref{Hseries}) become,
\begin{align}
&H_0=\int d^3x\Bigg[\frac{1}{2}(\pi_k)^2+\frac{1}{2}(\nabla\phi_k)^2
+\frac{m^2}{2}(\phi_k)^2\Bigg],\\
&H_1=\int d^3x\Bigg[\frac{\mu_{ijk}}{3!}\phi_i\phi_j\phi_k
+\delta_{t,k}\phi_k\Bigg],\\
&H_2=\int d^3x\Bigg[\frac{\lambda_{ijkl}}{4!}\phi_i\phi_j\phi_k\phi_l
+\frac{\delta_{m,jk}}{2}\phi_j\phi_k
+\frac{\delta_{Z,jk}}{2}\Big(\pi_j\pi_k-(\nabla\phi_j\cdot\nabla\phi_k)\Big)\Bigg],
\end{align}
where we introduced the counterterms (also symmetric under
permutations of indices) to absorb the divergences and use
more sensible renormalized perturbation theory.

Because of the way $H_2$ contributes to $h_2$ in (\ref{EquivH})
in the process up to the second order of $g^2$ the $\phi^4$ term
gives a contribution only as a local interaction term. Thus we
will set for the rest of the paper $\lambda_{ijkl}=0$. However
in higher orders one needs to take into account both the $\phi^4$ term
and its counterterm.

From now on we will represent all operators in the interaction picture.
We will use the following definition for the Feynman propagators and
related functions,
\begin{align}
D^{(jk)}_F(x-y)\equiv&
\langle 0|\mathrm{T}\{\phi_j(x)\phi_k(y)\}|0\rangle,\\
\overline{D}^{(jk)}_F(x-y)\equiv&
\langle 0|\overline{\mathrm{T}}\{\phi_j(x)\phi_k(y)]\}|0]\rangle,\\
D^{(jk)}(x-y)\equiv&
[\phi_j(x),\phi_k(y)]
=\varepsilon(x^0)\Big(D^{(jk)}_F(x)-\overline{D}^{(jk)}_F(x)\Big),\\
\widetilde{D}^{(jk)}(x-y)\equiv&
D^{(jk)}_F(x-y)+\overline{D}^{(jk)}_F(x-y),
\end{align}
where $\mathrm{T}$ and $\overline{\mathrm{T}}$ are chronological
and antichronological products respectively,
\begin{align}
&\mathrm{T}\{\phi_j(x)\phi_k(y)\}=
\theta(x^0-y^0)\phi_j(x)\phi_k(y)+
\theta(y^0-x^0)\phi_k(y)\phi_j(x),\\
&\overline{\mathrm{T}}\{\phi_j(x)\phi_k(y)\}=
\theta(x^0-y^0)\phi_k(y)\phi_j(x)+
\theta(y^0-x^0)\phi_j(x)\phi_k(y).
\end{align}
Because all these functions are proportional to the
$\delta_{jk}$ symbol we will
also use $D^{(k)}(x)\equiv D^{(kk)}(x)$. This yields the following
momentum representation,
\begin{align}
D(x-y)=\int\frac{d^4p}{(2\pi^4}D(p)e^{-ip(x-y)},\\
D^{(k)}_F(p)=\frac{i}{p^2-m_k^2+i\epsilon},
\quad
\overline{D}^{(k)}_F(p)=-\frac{i}{p^2-m_k^2-i\epsilon},\\
D^{(k)}(p)=\frac{2\pi}{2E^{(k)}_{\vec{p}}}
\Big(\delta(p^0-E^{(k)}_{\vec{p}})-\delta(p^0+E^{(k)}_{\vec{p}})\Big),
\label{Dfourier}
\end{align}
where $E^{(k)}_{\vec{p}}\equiv\sqrt{m_k^2+\vec{p}^2}$.

We now may consider the following correlation functions,
\begin{align}
\mathcal{G}_{H,n}(1,\ldots n)\equiv\langle\Omega|\mathrm{T}\{\phi_{i_1}(x_1)\ldots
\phi_{i_n}(x_n)\}|\Omega\rangle_H
=
\frac{\langle 0|\mathrm{T}\{\phi_{i_1}(x_1)\ldots
\phi_{i_n}(x_n)S_H\}|0\rangle}
{\langle 0|S_H|0\rangle}.
\end{align}
We assume that they can be obtained by simple analytical continuation
of the correlation functions in the Hermitian $\phi^3$ model to the
complex values of the coupling constant (this was argued to be valid in
\cite{BenderSD1} in contrast to the similar approach to the $-g^2\phi^4$
model plagued by breakdown of the ordinary perturbation theory).
It is interesting that because with extra assumption
(\ref{QextraT}) the metric operator $\eta^\dagger\eta$ happens to
act trivially on the interacting vacuum $|\Omega\rangle$ the inner product
in this correlation function may be understood as the modified inner product
$(\Psi,\Phi)$ and thus is conserved \cite{JonesCop}. This was used extensively
to study the path integral of the non-Hermitian model. However it is not clear
what relation these correlation functions have to observables as $\phi(x)$ are
not Hermitian with respect to the $(\Psi,\Phi)$. Also because $\phi(x)$ do
not commute with $\eta^\dagger\eta$ this trick is not particularly helpful for
the scattering processes.

The easiest counterterm to take into account is the tadpole counterterm.
It represents the shift of the field and in the Hermitian sigma-model it is
usually fixed so that the field was zero in the vacuum,
\begin{equation}
\langle\Omega|\phi(x)|\Omega\rangle_H=0,
\end{equation}
that yields that the tadpole absorbs the singular term without any extra
regular part,
\begin{equation}
\delta_{t,k}=-\frac{\mu_{kjj}}{2}D_F^{(j)}(0),
\label{tadpole}
\end{equation}
This may also be rewritten as the statement that the interaction Hamiltonian
is normally ordered,
\begin{equation}
H_1=\int d^3x\Bigg[:\frac{\mu_{ijk}}{3!}\phi_i\phi_j\phi_k:\Bigg],
\end{equation}
In the pseudo-Hermitian model we could add the real regular part to
$\delta_{t,k}$. Unlike the Hermitian case the shift of the field would be done
in the imaginary axis and would become a part of the intertwining operator $Q$.
This would be equivalent to certain shifts of the masses and coupling constants.
To not overcomplicate things we set tadpole counterterm to (\ref{tadpole}).

Then we renormalize the propagator,
\begin{equation}
\Pi^{(i_1i_2)}_H(p)=\int d^4(\Delta x) e^{ip\Delta x}\mathcal{G}_{H,2}(1,2),
\end{equation}
In the perturbation theory we can represent it as,
\begin{equation}
\Pi^{(i_1i_2)}_H\simeq D_F^{(i_1i_2)}
+ig^2D_F^{(i_1)}\Delta^{i_1i_2}_HD_F^{(i_2)},
\end{equation}
where,
\begin{equation}
\Delta^{i_1i_2}_H=-\frac{i}{2}\mu_{i_1jk}\mu_{i_2jk}
I_{jk}(p)+\delta_{m,i_1i_2}-\delta_{Z,i_1i_2}p^2,
\end{equation}
\begin{equation}
I_{jk}(p)=\int \frac{d^4k}{(2\pi)^4}D^{(j)}_F(k)D^{(k)}_F(p-k),
\label{loopintegral}
\end{equation}
Counterterms $\delta_{m,ij}$ and $\delta_{Z,ij}$ are used
to absorb the divergent part of this integral that
depends on the choice of the regularization, substraction scheme and
renormalization conditions. Our results do not depend on this choicess
but to satisfy the assumptions (\ref{Hseries}) we assume that
the counterterms are real.

\section{Correlation functions in the equivalent Hermitian QFT
\label{Correlators}}

We now are interested in computation of the correlation functions in the
equivalent Hermitian model using the standard formula,
\begin{align}
\mathcal{G}_n(1\ldots n)\equiv
\langle\Omega|\mathrm{T}\{\phi_{i_1}(x_1)\ldots
\phi_{i_n}(x_n)\}|\Omega\rangle
=
\frac{\langle 0|\mathrm{T}\{\phi_{i_1}(x_1)\ldots
\phi_{i_n}(x_n)S_h\}|0\rangle}
{\langle 0|S_h|0\rangle},
\end{align}
where $S_h$ is understood as in (\ref{Shmatrix}). To compute it we use
the Dyson series and the relation (\ref{EquivH}),
\begin{equation}
S_h\simeq 1-ig^2\int\limits_{-\infty}^{+\infty}dt\,
\Big[-H_2(t)+\frac{i}{2}[H_1(t),Q(t)]\Big]
\end{equation}
As $H_2(t)$ is the integral of the local operator we may compute its
contribution in a standard way. The real challenge is in the computation
of the contribution of the commutator. To compute it we remember
(\ref{Qformula}) to represent it as,
\begin{equation}
[H_1(t),Q(t)]=\partial_{\tau}^{-1}[H_1(t),H_1(\tau)]\Big\vert_{\tau=t}
\end{equation}
where we define the antiderivative according to our choices
in Section \ref{QinQFT},
\begin{equation}
\partial_{\tau}^{-1}e^{iEt}\equiv
-ie^{iEt}\left[
\mathrm{P.v.}\frac{1}{E}
+i\alpha\delta(E)\right].
\label{antiderivative}
\end{equation}
As in most practical QFT computations we will assume that all our operators
are sufficiently nice operator valued distributions and this operation
commutes with all integrals except integrals in $t$ and $\tau$. Also we will
assume that it commutes with the operation of taking the vacuum
expectation value.

Taking the local operator representation of $H_1(t)$,
\begin{equation}
H_1(t)=\int d^3w V(w),\quad w^0\equiv w
\end{equation}
we represent the contribution of the commutator to the S-matrix as,
\begin{align}
\delta S_h=-ig^2\int\limits_{-\infty}^{+\infty}dt\,
\frac{i}{2}[H_1(t),Q(t)]
=-ig^2\int d^4z d^4w \delta(w^0-z^0)\partial_{w^0}^{-1}
\mathcal{W}(z,w),
\end{align}
where,
\begin{equation}
\mathcal{W}(z,w)=\frac{i}{2}[V(z),V(w)]
\end{equation}

Finally with all our assumptions we write the contibution to the
correlation function as,
\begin{align}
\delta\mathcal{G}_n(1\ldots n)
=-ig^2\int d^4z\, d^4w \delta(w^0-z^0)
\partial_{w^0}^{-1}\mathcal{F}_n\Big(1\ldots n|z,w\Big)
\end{align}
where we introduced,
\begin{align}
\mathcal{F}_n\Big(1\ldots n|z,w\Big)
\equiv&\langle 0|\mathrm{T}\{\phi_{i_1}(x_1)\ldots\phi_{i_n}(x_n)
\mathcal{W}(z,w)\}|0\rangle\nonumber\\&
-\langle 0|\mathrm{T}\{\phi_{i_1}(x_1)\ldots\phi_{i_n}(x_n)\}
|0\rangle\langle 0|\mathcal{W}(z,w)\}|0\rangle
\end{align}

To use that formula all we need is to represent $\mathcal{W}(z,w)$
in terms of the chronological product. To do so we first rewrite it as,
\begin{align}
\mathcal{W}(z,w)=\frac{i}{2}\varepsilon(z^0-w^0)
\cdot\Big[\mathrm{T}\{V(z)V(w)\}-
\overline{\mathrm{T}}\{V(z)V(w)\}]\Big],\label{Wcom1}
\end{align}

We will consider the model from the previous section, i.e.,
\begin{equation}
V(x)=:\frac{\mu_{ijk}}{3!}\phi_i(x)\phi_j(x)\phi_k(x):
\end{equation}

Then we use the Wick's theorem for chronological product and similar
statement for the antichronological product (that works by simply
replacing $D_F\mapsto \overline{D}_F$) to convert (\ref{Wcom1})
to the normal form. The normal ordering of $V$ means that we
should not include the terms with contractions of fields with same
coordinates. As result we get,
\begin{align}
\mathcal{W}(z,w)=\frac{i}{8}\mu_{ijk}\mu_{lmk}\Bigg[&
:\phi_i(z)\phi_j(z)\phi_l(w)\phi_m(w):D^{(k)}(z-w)\nonumber\\&
+2:\phi_i(z)\phi_l(w):D^{(k)}(z-w)\widetilde{D}^{(jm)}(z-w)
+\Big(\text{$c$-number}\Big)\Bigg],
\end{align}
We do not track the $c$-number as its contribution is canceled out
by the vacuum normalization. Then we use the Wick's theorem again
to rewrite it as,
\begin{align}
\mathcal{W}(z,w)=\frac{i}{8}\mu_{ijk}\mu_{lmk}\Bigg[&
+\mathrm{T}\{:\phi_i(z)\phi_j(z)::\phi_l(w)\phi_m(w):\}D^{(k)}(z-w)\nonumber\\
&-4\mathrm{T}\{\phi_i(z)\phi_l(w)\}D^{(k)}(z-w)D^{(jm)}_F(z-w)\nonumber\\
&+2\mathrm{T}\{\phi_i(z)\phi_l(w)\}D^{(k)}(z-w)\widetilde{D}^{(jm)}(z-w)
+\Big(\text{$c$-number}\Big)\Bigg],
\label{Wformula}
\end{align}

\section{Hermitian field propagator\label{Propagator}}

The propagator is computed according to the standard formula,
\begin{equation}
\Pi^{(i_1i_2)}(p)=\int d^4(\Delta x) e^{ip\Delta x}\mathcal{G}_2(1,2),
\end{equation}

For the 2-point contribution of $\mathcal{W}$ we obtain,
\begin{align}
&\mathcal{F}(1,2|z,w)=
\frac{i}{4}\mu_{i_1jk}\mu_{i_2jk}
D^{(k)}(z-w)\widetilde{D}_F^{(j)}(z-w)
\nonumber\\&
\Bigg[D_F^{(i_1)}(x_1-z)D_F^{(i_2)}(x_2-w)
+D_F^{(i_1)}(x_1-w)D_F^{(i_2)}(x_2-z)\Bigg].
\label{F2}
\end{align}
Using our choice of (\ref{antiderivative}) we find that the first term
gives the following contribution to the propagator,
\begin{align}
&\delta\Pi^{(i_1i_2)}_1(p)=-\frac{ig^2}{4}\mu_{i_1jk}\mu_{i_2jk}
\int d^4(\Delta x)\int dz^0\int d^3z d^3w
\int \frac{d^4p_1}{(2\pi)^4}\frac{d^4p_2}{(2\pi)^4}
\frac{d^4q}{(2\pi)^4}\frac{d^4r}{(2\pi)^4}
\nonumber\\&
e^{i(p-p_1)\Delta x-i(p_1+p_2)x_2}
e^{i(p_1^0+p_2^0)z^0-i\vec{p}_1\vec{z}-i\vec{p}_2\vec{w}
+i(\vec{q}+\vec{r})(\vec{z}-\vec{w})}
D_F^{(i_1)}(p_1)D_F^{(i_2)}(p_2)
\widetilde{D}_F^{(j)}(q)D_F^{(k)}(r)
\nonumber\\&
\Bigg[\mathrm{P.v.}\frac{1}{q^0+l^0+p_2^0}
+i\alpha\delta(q^0+l^0+p_2^0)\Bigg],
\end{align}
Integrals over $\Delta x$, $z^0$, $\vec{z}$ and $\vec{w}$ introduce
the $\delta$-functions that fix the momenta, and using the momentum
representation of $D^{(k)}$ in (\ref{Dfourier}) we obtain,
\begin{align}
&\delta\Pi^{(i_1i_2)}_1(p)=-\frac{ig^2}{4}\mu_{i_1jk}\mu_{i_2jk}
D_F^{(i_1)}(p)D_F^{(i_2)}(p)
\int\frac{d^4q}{(2\pi)^4}
\widetilde{D}_F^{(j)}(q)
\Bigg[-\mathrm{P.v}\frac{1}{(p-q)^2-m_k^2}
\nonumber\\&
+i\alpha\varepsilon(p^0-q^0)\delta\Big((p-q)^2-m_k^2\Big)\Bigg].
\end{align}
Similar computation for the contribution of the second term in (\ref{F2})
yields,
\begin{align}
&\delta\Pi^{(i_1i_2)}_2(p)=-\frac{ig^2}{4}\mu_{i_1jk}\mu_{i_2jk}
D_F^{(i_1)}(p)D_F^{(i_2)}(p)
\int\frac{d^4q}{(2\pi)^4}
\widetilde{D}_F^{(j)}(q)
\Bigg[-\mathrm{P.v}\frac{1}{(p-q)^2-m_k^2}
\nonumber\\&
-i\alpha\varepsilon(p^0-q^0)\delta\Big((q-p)^2-m_k^2\Big)\Bigg].
\end{align}
Thus the $\alpha$-term cancels out. As the counterterms give the
standard contribution the full propagator can be written in the standard
way,
\begin{equation}
\Pi^{(i_1i_2)}\simeq D_F^{(i_1i_2)}
+ig^2D_F^{(i_1)}\Delta^{(i_1i_2)}D_F^{(i_2)},
\end{equation}
where,
\begin{align}
&\Delta^{(i_1i_2)}=\frac{1}{2}\mu_{i_1jk}\mu_{i_2jk}
\Bigg(\Im\Big[I_{jk}(p)\Big]+
\Im\Big[\widetilde{I}_{jk}(p)\Big]\Bigg)
+\delta_{m,i_1i_2}-\delta_{Z,i_1i_2}p^2,
\end{align}
where $I_{jk}(p)$ is given by (\ref{loopintegral}) and,
\begin{align}
&\widetilde{I}_{jk}(p)
=-\int\frac{d^4q}{(2\pi)^4}
\frac{1}{q^2-m_j^2+i\epsilon}
\frac{1}{(p-q)^2-m_k^2-i\epsilon}
\end{align}
This loop integral is problematic as the standard Wick
rotation fails and its accurate computation may significantly
depend on the regularization. The issues with the Wick rotation
issues are characteristic for the nontrivial QFT setting e.g. on curved backgrounds
or in the noncommutative geometry models \cite{VisserWick,DAndreaWick} However we present
a simple symmetry argument that hopefully allows us to
omit this integral altogether. We note that under $j\leftrightarrow k$
these integrals transform as,
\begin{equation}
I_{jk}(p)=I_{kj}(p),\quad
\widetilde{I}_{jk}(p)=\widetilde{I}_{kj}^\ast.
\end{equation}
Therefore $\Im\Big[I_{jk}(p)\Big]$ is symmetric while
$\Im\Big[\widetilde{I}_{jk}(p)\Big]$ is antisymmetric. Therefore
the latter does not make any contribution. Thus,
\begin{align}
&\Delta^{(i_1i_2)}=\frac{1}{2}\mu_{i_1jk}\mu_{i_2jk}
\Im\Big[I_{jk}(p)\Big]+
+\delta_{m,i_1i_2}-\delta_{Z,i_1i_2}p^2=
\Re\Big[\Delta^{(i_1i_2)}_H\Big],
\end{align}
where in the end we used the reality of the counterterms.
This is consistent with our general formula (\ref{FinalT}).
The most interesting aspect of this result is that if
non-Hermitian model admitted two body,
some of $\Delta^{(i_1i_2)}_H$ would get the
imaginary part corresponding to the decay width. However
because $h_1=0$ the equivalent Hermitian model never
admits any two body decays. Thus according to the optical
theorem to keep unitarity $\Delta^{(i_1i_2)}$ must remain real
and indeed this is what happens.

\section{2 to 2 scattering amplitude\label{ScatteringAmplitude}}

To find 2 to 2 scattering amplitude we first separate the connected part of the
4-point correlation function,
\begin{align}
&\mathcal{G}_4(1,2,3,4)=\mathcal{G}_{4,c}(1,2,3,4)
+\mathcal{G}_{2}(1,2)\mathcal{G}_{2}(3,4)
+\mathcal{G}_{2}(1,3)\mathcal{G}_{2}(2,4)
+\mathcal{G}_{2}(1,4)\mathcal{G}_{2}(2,3).
\end{align}
To compute the $\mathcal{G}_{4,c}$ one needs to include only the terms
where all external fields $\phi_{i_k}(x_k)$ are contracted with fields in
$\mathcal{W}$. That means that only the only the 4-field term in (\ref{Wformula})
gives contribution. It can be further separated into three scattering channels,
\begin{align}
&\mathcal{G}_{4,c}(1,2,3,4)
=\widetilde{\mathcal{G}}_{4}(1,2|3,4)
+\widetilde{\mathcal{G}}_{4}(1,3|2,4)
+\widetilde{\mathcal{G}}_{4}(1,4|2,3),
\end{align}
\begin{align}
&\widetilde{\mathcal{G}}_{4,c}(1,2|3,4)
=-ig^2\int d^4z\, d^4w \delta(w^0-z^0)
\partial_{w^0}^{-1}
\widetilde{\mathcal{F}}_{4}(1,2|3,4|z,w)
\end{align}

For each scattering channel we get,
\begin{align}
&\widetilde{\mathcal{F}}_{4}(1,2|3,4|z,w)
=\frac{i}{2}\mu_{i_1i_2k}\mu_{i_3i_4k}D^{(k)}(z-w)\Bigg[
\nonumber\\&
D_F^{(i_1)}(x_1-z)D_F^{(i_2)}(x_2-z)
D_F^{(i_3)}(x_3-w)D_F^{(i_4)}(x_4-w)
+\Big((1,2)\leftrightarrow (3,4)\Big)\Bigg].
\end{align}

In the momentum representation,
\begin{align}
&\widehat{\widetilde{\mathcal{G}}}_{4,c}(1,2|3,4)
\equiv \int \prod_k\Big\{d^4x_k\,e^{ip_kx_k}\Big\}
\widetilde{\mathcal{G}}_{4,c}(1,2|3,4)
\nonumber\\&
=\frac{g^2}{2}\mu_{i_1i_2k}\mu_{i_3i_4k}
D_F^{i_1}(p_1)D_F^{i_2}(p_3)D_F^{i_4}(p_3)D_F^{i_4}(p_4)
\nonumber\\&
\Bigg[A^{(k)}(p_1+p_2|p_3+p_4)+A^{(k)}(p_3+p_4|p_1+p_2)\Bigg],
\end{align}
where we introduced,
\begin{align}
&A^{(k)}(p_{in}|p_{out})=
-i\int  dz^0\,d^3z d^3w
\int\frac{d^4q}{(2\pi)^4}D^{(k)}(q)
e^{i(p_{in}^0+p_{out}^0)z^0+i\vec{q}(\vec{z}-\vec{w})
-i\vec{p}_{in}\vec{z}-i\vec{p}_{out}\vec{w}}
\nonumber\\&
\Bigg[\mathrm{P.v.}\frac{1}{p_{out}^0-q^0}
+i\alpha\delta(p_{out}^0-q^0)\Bigg].
\end{align}
The integrals over $z$ and $w$ result in $\delta$-functions for momenta,
and using the momentum representation of $D^{(k)}$ in (\ref{Dfourier}) we obtain,
\begin{align}
&A^{(k)}(p_{in}|p_{out})
=-i(2\pi)^4\delta^{(4)}(p_{in}+p_{out})
\nonumber\\&
\Bigg[\mathrm{P.v.}\frac{1}{(p_{out})^2-m_k^2}
+i\alpha\varepsilon(p_{out}^0)
\delta\Big((p_{out})^2-m_k^2\Big)\Bigg]
\end{align}

Because the $\alpha$-term happens to be antisymmetric
under $p_{in}\leftrightarrow p_{out}$ exchange it cancels out and
we get,
\begin{align}
&\widehat{\widetilde{\mathcal{G}}}_{4,c}(1,2|3,4)
=-i(2\pi)^4\delta^{(4)}\Big(\sum_n p_n\Big)
g^2\mu_{i_1i_2k}\mu_{i_3i_4k}
\nonumber\\&
D_F^{i_1}(p_1)D_F^{i_2}(p_2)D_F^{i_3}(p_3)D_F^{i_4}(p_4)
\mathrm{P.v.}\frac{1}{(p_3+p_4)^2-m_k^2},
\end{align}
Let us apply the standard LSZ formula for the 2 to 2 scattering
amplitude,
\begin{align}
&\widehat{\mathcal{G}}_{4,c}(1,2,3,4)
\underset{\text{on-shell}}{\sim}
(2\pi)^4\delta^{(4)}(p_1+p_2+p_1+p_2)
\nonumber\\&
D_F^{(i_1)}(p_1)D_F^{(i_2)}(p_2)D_F^{(i_3)}(q_1)D_F^{(i_4)}(q_2)
\mathcal{A}_{2\to 2}^{(i_1i_2|i_3i_4)}(p_1,p_2|-p_3,-p_4),
\end{align}
we obtain a simple result,
\begin{align}
&\mathcal{A}_{2\to 2}^{(i_1i_2|j_1j_2)}(p_1,p_2|q_1,q_2)=
-ig^2\mu_{i_1i_2k}\mu_{j_1j_2k}
\Bigg[\mathrm{P.v.}\frac{1}{s-m_k^2}
+\mathrm{P.v.}\frac{1}{t-m_k^2}+\mathrm{P.v.}\frac{1}{u-m_k^2}\Bigg],
\label{2to2amplitude}
\end{align}
where $(s,t,u)$ are the standard Mandelstam variables.
One may recognize the scattering amplitude of the $\phi^3$ theory analytically
continued to the imaginary coupling. However the poles are taken in principal
value. This is in a perfect agreement with the general formula (\ref{FinalT}).

However this simple result has a significant problem.
The scattering amplitude (\ref{2to2amplitude}) violates the
well-known causality constraints on the analytical structure of the scattering
amplitude \cite{Eden,Sudarshan2}.
In fact we can represent the amplitude above as,
\begin{equation}
\mathcal{A}_{2\to 2}=-\frac{1}{2}\mathcal{A}_{2\to 2}^{(+i\epsilon)}
+\frac{1}{2}\mathcal{A}_{2\to 2}^{(-i\epsilon)}
\end{equation}
The first term corresponds to the 2 to 2 scattering amplitude in the
Hermitian $\phi^3$ QFT that respects the causal nature of the scattering
 - the scattered particles are produced only
in the future lightcone of the collision event of the ingoing particles. On the
hand the second term respects the reversed causality - the scattered particles are produced
only in the past lightcone of the collision event of the ingoing particles. Thus the total
scattering amplitude is acausal in a full accordance with (\ref{CausalityViolation}).
One could conjecture that this happens because of the extra restriction (\ref{QextraT})
and may be avoided by $\mathcal{T}$-violating extra term that
shifts the poles in (\ref{antiderivative}) to the lower complex half-plane.
However as we have shown the result happened to not depend on
$\alpha$. Indeed among the extra assumptions 
in our derivation of (\ref{FinalT}) we used only (\ref{Qextra}) but not
(\ref{QextraT}).

\section{Attempt at relaxing the Hermiticity assumption\label{NoHermiticity}}

One may conjecture that the equivalent Hermitian theory may
actually be causal but the field operators obtained with use of the
Hermitian intertwining operator correspond to nonlocal configurations.
Then there may exist an unitary transformation that makes the causal
nature of the model apparent. While not proving that such operator
does not exist we show how the simplest ansatz of this sort fails.

Let us take the Hermitian $Q$ satisfying (\ref{Qformula}) but
modify the formula for the intertwining operator,
\begin{equation}
\eta=e^{-g(1+i\theta)Q}+\mathcal{O}(g^3).
\end{equation}
Then the metric operator $\eta^\dagger\eta$ remains to be the same
and therefore such intertwining operator converts non-Hermitian theory
into the Hermitian one.

Repeating the computations from Section \ref{EtaPerturb} and
Section \ref{FormalS} one obtains the following equivalent
Hermitian Hamiltonian,
\begin{equation}
h^{(\theta)}\simeq H_0+g\theta H_1+g^2\Big(-H_2+\frac{1+\theta^2}{2}[H_1,Q]\Big)
\end{equation}
and the corresponding formal S-matrix looks like,
\begin{equation}
S_h^{(\theta)}\simeq 1+ig\theta T_H^{(1)}
+ig^2\Big(-T_H^{(2)}+\frac{i}{2}(1+\theta^2)[T_H^{(1)}]^2\Big)
\end{equation}
Using the results of the previous section it is trivial to confirm that the
2 to 2 scattering amplitude computed for $h^{(\theta)}$ is in agreement
with the above formula for $S_h^{(theta)}$ and can be written as,
\begin{align}
&\mathcal{A}_{2\to 2}^{(i_1i_2|j_1j_2)}(p_1,p_2|q_1,q_2)=
-ig^2\mu_{i_1i_2k}\mu_{j_1j_2k}
\Bigg[\mathrm{P.v.}\frac{1}{s-m_k^2}+i\pi\theta\delta(s-m_k^2)
\Bigg]\nonumber\\
&+\Big(\text{$t$ and $u$ terms}\Big)
\end{align}
Sadly $\theta$ pushes pole in the wrong direction. The most curious case
is when $\theta=1$. Then the first order $T$-matrix coincides
with a first order $T$-matrix of the \textit{Hermitian} $\phi^3$
model. However while the propagator remains to have the right $i\epsilon$
prescription the 2 to 2 scattering amplitude is analytical in the wrong complex
half-plane. Thus this simple ansatz can't help us restore causality.

\section{Conclusions\label{Conclusions}}

In this work we have considered the perturbative scattering in the local
non-Hermitian $\mathcal{PT}$-symmetric quantum field theory interpreted
in a pseudo-Hermitian fashion. We explicitly showed that the intertwining
operator remains to be non-trivial even when the interactions asymptotically
vanishes. In fact the intertwining operator in the limit of large times is
proportional to the first order T-matrix of the non-Hermitian model.
This makes it quite hard to associate the field variables of the initial
non-Hermitian model with any actual observables even in the asymptotic
region. Thus we resort to the computations with equivalent Hermitian Hamiltonian.

Despite this equivalent Hamiltonian being quite complicated and nonlocal
the surprising result is that the leading order of its S-matrix is very simply
related to the S-matrix of the original non-Hermitian model.
The generic consequences of this relation include the disappearance of
two body decay that are also reflected in the 1-loop correction to the mass
of particles but also the violation of the causal analytic structure of the
2 to 2 body scattering amplitude. These effects were explicitly demonstrated
for the $i\phi^3$ model. Our result raise a question whether the causality
violation in such models may prevent possible applications from being
phenomenologically viable.

The modifications of the analytic structure of the propagators that
destory the ordinary notion of the microscopic causality
were studied before in the context of the models with indefinite metrics
or $\mathcal{CPT}$-violation \cite{Nakanishi,KurkovCPT,ChaichianCPT}.
As a matter of fact our results bear strong similarity to the models
of the so-called shadow states \cite{Sudarshan1,Sudarshan2}.
While nowhere in our consideration the indefinite
norm appears (see \cite{MostafazadehIndef} and references therein for
the discussion on the positivity of the norm in the $\mathcal{PT}$-symmetric
models) one may conjecture that the pseudo-Hermitian interpretation of the local
$\mathcal{PT}$-symmetric QFT may be equivalent to the positive norm sector
of some QFT with indefinite metrics. At this level this conjecture remains to be
purely hypothetical and will be explored in the future work.

As was noted in \cite{ZnojilScatter1,ZnojilScatter2,ZnojilScatter3} the causality
issues of the finite dimensional $\mathcal{PT}$-symmetric quantum mechanics
may be tackled by introducing certain degree of nonlocality into the
non-Hermitian Hamiltonian. Similar strategy may work in case of QFT and the
simple form of our results may imply that certain modified notion of locality of
the non-Hermitian Hamiltonian may arise.

In the end of Section \ref{SinQFT} we mentioned that the case of the linear
potential is exceptional and
preserves causality. This happens because the intertwining operator becomes
simply a shift of the field variable in the imaginary direction and thus
preserves the local structure of the Hamiltonian. Because we restricted
ourselves to the lowest orders of perturbation theory about free bosonic QFT
with polynomial interactions this case may appear to be somewhat trivial.
However in \cite{BenderDual} it was shown that the shift intertwining operator
can be used for the nonperturbative construction of the
$\mathcal{PT}$-symmetric Sine-Gordon model.

We conclude by reviewing several shortcomings one may see in our
consideration.
\begin{itemize}

\item The results presented in this paper are derived with a level of rigor typical
for the practical QFT computations that involves rather liberal operations
with integrals of the operator-valued distributions and no disscussions of
the domain issues and existence of operators in question beyond the
perturbation theory level. It should be stressed however that this sloppy
(though highly fruitful) approach is characteristic also to the field of the
$\mathcal{PT}$-symmetric quantum theories even when finite-dimensional
quantum mechanics is considered. In opinion of the author many questions
about the Hilbert spaces involved and existence of the intertwining operator
in a strict sense (not just on the Hamiltonian eigenstates but also on a generic
quantum state) are yet to be addressed. Thus we deem it premature to
consider the $\mathcal{PT}$-symmetric QFT on a more rigorous level and
are prepared for the loopholes associated with singular transformations of
a generic wave functional even within the realm of the perturbation theory.

\item As we mentioned in Section \ref{NoHermiticity} the causality problem
may be simply an artifact of choosing the bad variables and some unitary
transformation may fix it. We were not able to do so by choosing the simplest
ansatz but it may be possible to find the more appropriate one by studying
accurately the asymptotic conditions. Our resuls were also restricted to the
lowest nontrivial order with computations in the higher orders being
much more nontrivial.

\item Throughout the paper we assumed the applicability of the perturbation
theory as it is presented in Section \ref{EtaPerturb}. The acausal poles may
actually signify its breakdown and may disappear after an appropriate
resummation. The applicability of the perturbation theory
for the formal correlation functions in the $\mathcal{PT}$-symmetric
non-Hermitian QFT was studied in \cite{BenderSD1}. But again for 
the correlation functions of the equivalent Hermitian model we have no simple
form beyond the lowest order and thus leave this important question for the
future investigation.

\item As we mentioned above the solution (\ref{Qfinal}) for the intertwining
operator that has smooth
momentum representation and does not break the Lorentz invariance
is not unique because an arbitrary integral of motion of the free QFT can be
added. This may greatly improve the locality of the resulting equivalent
Hermitian Hamiltonian
\footnote{The author would like to thank the referee for this observation and
pointing out an example.}. Our general formula (\ref{Tmatrix1}) implies that
for this to work the commutator $[Q_{out},Q_{in}]$ must not vanish.
One may give a specific
example in case of the fermionic QFT considered in
\cite{BenderDual}. In that case one may construct the quadratic
non-Hermitian
interaction satisfying the parity conventions (\ref{Hseries}),
\begin{equation}
\mathcal{L}=\bar{\psi}(i\gamma^\mu\partial_\mu-m_1-m_2\gamma_5)\psi,
\end{equation}
Interestingly the aforementioned paper present the local intertwining operator
in the form,
\begin{equation}
Q\sim \int dx\,\psi^\dagger\gamma_5\psi,
\end{equation}
whereas the ansatz similar to (\ref{Qfinal}) would produce a different highly
nonlocal intertwining operator,
\begin{equation}
Q\sim \int dx\,\bar{\psi}\Big(-i\gamma^k\partial_k+m\Big)^{-1}\psi
\end{equation}

However one may see many differences with the
bosonic case considered in
this paper. With bosonic fields $H_1$ should contain the odd number of fields
to have the appropriate $\mathcal{P}$ parity. Moreso the spinor field rotation
does not introduce derivative terms into the interaction Hamiltonian because
$\Psi^\dagger$ serves as a canonical momentum for $\Psi$ as opposed to
$\dot{\phi}$ in the bosonic case. Thus the fermionic $\mathcal{PT}$-symmetric
QFT may be more well behaved and deserves additional analysis. At this
point it is not clear whether good $Q$ that preserves both causality and
Lorentz invariance exists for bosonic field interactions.

\end{itemize}

\begin{acknowledgments}
The author would like to thank A.A. Andrianov, M.V. Ioffe, A.V. Golovnev and
M.A. Kurkov for helpful discussions and support. The funding for this work
was provided by the RFBR project 18-02-00264.
\end{acknowledgments}


\providecommand{\noopsort}[1]{}\providecommand{\singleletter}[1]{#1}%

\end{document}